**Hiroyuki Yamamoto**・**Kentaro Abe**・**Yoshiharu Arakawa**・**Takashi Okuyama**・**Joseph Gril**

# Role of the Gelatinous Layer (G-Layer) on the Origin of the Physical Properties of the Tension Wood of *Acer sieboldianum*

**Abstract**   The tension wood (TW) properties of a 70 year-old *Acer sieboldianum* Miq were analyzed by using the G-fiber model which was proposed in our previous paper. The roles of the G-layer on the origins of (1) a high large tensile growth stress, (2) a large longitudinal Young's modulus, and (3) a high longitudinal drying shrinkage in the tension wood xylem were discussed on the basis of the simulations using the G-fiber model. The results suggest that the G-layer generates a high tensile stress in the longitudinal direction during the xylem maturation; the longitudinal Young's modulus of the green G-layer becomes significantly higher than that of the lignified layer; furthermore, the G-layer tends to shrink extraordinarily higher than that of the lignified layer during the moisture desorption.

**Key words**   Cellulose microfibril・Gelatinous fiber・Growth stress・wood cell wall・Reaction wood

Hiroyuki Yamamoto *・Kentaro Abe・Yoshiharu Arakawa・Takashi Okuyama
(*) School o Bioagricultural Sciences, Nagoya University, Chikusa, Nagoya 464-8601 Japan
Tel. +52-789-4152; Fax +81-52-789-4150,   E-mail: hiro@agr.nagoya-u.ac.jp

Joseph Gril
Laboratoire de Mecanique et Genie Civil, Universite Montpellier 2, Place E. Bataillon, cc048, 34095 Montpellier France

## Introduction

Tension wood (TW) shows abnormal xylem properties as compared to the normal wood (NW), e.g. a large tensile growth stress, a high longitudinal Young's modulus, and a large longitudinal shrinkage after drying. Some researchers attribute those behaviors to the physical properties of the gelatinous layer (G-layer) through comparing xylem properties and anatomical features in the TW xylem.[1-4]

To verify their ideas, it is required to observe the behaviors or physical properties of the G-layer by isolating it directly from the lignified layer. However, it is almost impossible to obtain the G-layer cylinder without giving any damages. No matter how we could obtain an isolated G-layer cylinder, it is still difficult to provide an accurate measurement since the isolated G-layer cylinder is too small to be analyzed by the ordinary mechanical testing machine.

The authors consider that a simulation using a mathematical model of the multi-layered wood fiber gives one of the most effective approaches for estimating the behavior of each cell wall constituent as it is in the cell wall.[5] In the previous report, we proposed a structural model of the G-fiber consisting of four-layered cylinders (CML+S1+S2+G), and formulated the mechanical behaviors of the G-fiber model on the basis of the reinforced-matrix hypothesis.[6]

The formula derived in the previous report contains several parameters. We need to optimize those parameters so as to obtain very reasonable result when we simulate the observed phenomena on the basis of the G-fiber model. Conversely to say, it can be considered that the optimized values of the parameters should reflect certain internal properties and fine composite structures of each constituent material in the G-layer. In this report, based on the simulation using the G-fiber model, we analyzed the observed results on the physical properties of the TW xylem which was formed in an inclined stem of a 70-years-old Kohauchiwakaede (*Acer sieboldianum* Miq.), and we tried to explain the role of the G-layer on the origin of distinctive xylem properties in the TW.

## Experiment [3]

Material and method

A 70-year-old kohauchiwakaede (*Acer sieboldianum* Miq.), grown on a steep slope at a private mountain in Kiyomi-cho, Gifu prefecture, Japan, 14 cm in DBH, having a leaning stem, was used for the experiment At the breast height, ten measuring points were set peripherally on the xylem surface of the leaning stem. At each point, the released strain of the longitudinal growth stress on the xylem surface ($\varepsilon_L^X$) was measured by using the ordinary strain-gauge method in early April 1988. Thereafter, rectangular portions, 70×10×5 mm and 50×10×5 mm in the longitudinal (L), the tangential (T), and the radial (R) directions, respectively, were sampled away from the upper or lower positions at each measuring point of the released strain. Then, respective portions were used for obtaining the tensile Young's modulus under the green condition ($E_L^X$) and the longitudinal shrinkage ($\alpha_L^X$) from green to oven-dried condition.[3]

After that, transverse section, 10μm in thickness, was cut from each measuring point of the released strain by the sliding microtome, and it was stained by safranin and ferric hematoxylin, thereafter, it was mounted on a slide glass with the jelly-like compound of gelatin, glycerin, and water. By using the light microscope connected to the image processor, microscopic images at the large and small magnification were photographed within the outermost annual ring of the mounted section. From the images photographed at the small magnification, the area composition of domain of each tissue, e.g. vessel element ($V$), ray tissue ($R$), and wood fiber ($F$), was computed. From the images at the large magnification, the area ratios of the lignified layer ($s$), the G-layer ($g$), and the cell lumen in the domain of the wood fiber were determined. Frequency of the G-fiber per unit area ($N_g$) in the domain of the wood fiber and that of the normal wood fiber (N-fiber) ($N_n$) were also counted.

Flat-sliced samples, 5×5×0.015 mm in L, T, R respective directions, were cut from the outermost annual rings of both the NW xylem and the highly-developed TW xylem. Sampled specimens from the TW xylem were quickly dried with ethanol, and were treated with an ultra-sonic vibrator in water to remove the G-layers from the lignified layer.[7] Thereafter, the microfibril angles in the middle layer of the secondary wall (MFA) were measured by the iodine-staining method.[8]

Observed Results

Obtained results were overviewed in Table 1, which was already reported in our previous paper.[3] From the Table 1, it can be clearly understood that the TW xylem shows quite distinctive properties as compared to the NW xylem. It is considered that either of the G-layer formation or the relatively small MFA in the S2 layer of the G-fiber would cause the distinctive xylem properties in the TW. However, it is still unsolved which possibility is more positively concerned with the origin of the TW properties, or there is something other factor which causes the TW properties. In the present paper, we tried to answer this question through simulating the mechanical behaviors of the G-fiber on the basis of the formula derived in our previous paper.[6]

Table 1. Observed data in a 70-year-old Kohauchiwakaede (*Acer sieboldiunum* Miq.).

| Measuring positions | 1 | 2 | 3 | 4 | 5 | 6 | 7 | 8 | 9 | 10 |
|---|---|---|---|---|---|---|---|---|---|---|
| [Area composition of domain of each tissue (%) - measured at the small maginification ] | | | | | | | | | | |
| Vessel [V] | 7.2 | 6.2 | 7.3 | 7.7 | 8.8 | 8.0 | 9.0 | 7.9 | 5.1 | 9.9 |
| Ray [R] | 13.0 | 17.8 | 12.8 | 12.6 | 12.3 | 17.6 | 14.7 | 12.3 | 16.6 | 10.8 |
| Fiber [F] | 79.8 | 76.0 | 79.9 | 79.7 | 78.9 | 74.4 | 76.3 | 79.8 | 78.3 | 79.3 |
| [Area ratios of the wall in the wood fiber domain (%) - measured at the large maginification ] | | | | | | | | | | |
| G-layer [g] | 8.77 | 16.4 | 7.88 | 10.20 | 0.507 | 1.61 | 0 | 0 | 0 | 0 |
| Lignified layer [s] | 70.2 | 60.5 | 63.2 | 62.5 | 69.1 | 63.8 | 60.6 | 62.0 | 67.9 | 65.9 |
| [Frequency of the fiber cell in the wood fiber domain (/ 0.01 mm$^2$) ] | | | | | | | | | | |
| Total wood fiber | 99 | 107 | 98 | 102 | 85 | 95 | 120 | 84 | 86 | 82 |
| ( G-fiber ) | (51) | (73) | (34) | (44) | (3) | (11) | ( 0 ) | ( 0 ) | ( 0 ) | ( 0 ) |
| Relative frequency of G-fiber [ φ ] | 0.51 | 0.68 | 0.35 | 0.43 | 0.04 | 0.12 | 0 | 0 | 0 | 0 |
| [Physical properties ] | | | | | | | | | | |
| Released strain [$\varepsilon_L^X$] (%) | -0.218 | -0.204 | -0.094 | -0.190 | -0.024 | -0.038 | -0.003 | -0.039 | 0.004 | -0.020 |
| Oven-dried shrinkage [$\alpha_L^X$] (%) | 1.16 | 1.10 | 0.48 | 0.93 | 0.24 | 0.52 | 0.51 | 0.50 | 0.22 | 0.018 |
| Green Young's modulus [$E_L^X$] (GPa) | 10.78 | 9.21 | 8.42 | 11.07 | 8.92 | 6.57 | 7.15 | 6.37 | 5.59 | 7.35 |
| [Microfibril angle in the S2 layer (deg.) ] | **Tension wood region** | | | | | **Normal wood region** | | | | |
| Number of the measured fiber | 32 | | | | | 32 | | | | |
| Average [ θ ] (deg) | 23.1 | | | | | 27.2 | | | | |
| Standard deviation (deg) | (2.18) | | | | | (3.38) | | | | |

## Simulation

### G-fiber model

A schematic model of the typical G-fiber, consisting of the compound middle lamella (CML), the S1, the S2, and the G-layers, was shown in Fig.1.[6]

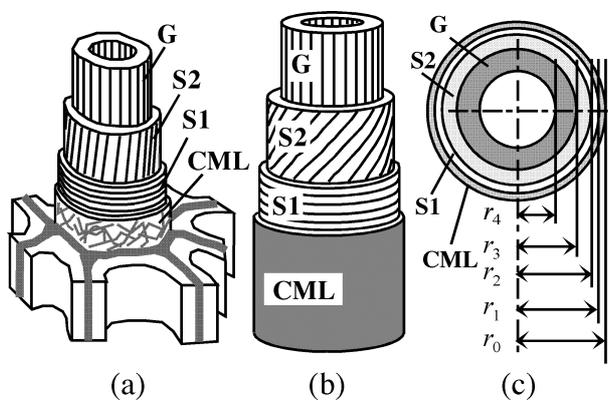

Fig.1. Multilayered structure of the gelatinous fiber. (a); Microscopic structure, (b); A mechanical model, (c); Crosscut surface of the mechanical model. Each consists of compound middle lamella (CML), outermost layer of the secondary wall (S1), its middle layer (S2), and gelatinous layer (G).

(a)   (b)   (c)

### Parameters in the basic formula

In this report, we focused on three biomechanical processes in the TW xylem, e.g. (1) cell wall maturation, (2) elastic deformation due to action of an axial traction under the moisture steady condition, and (3) moisture adsorption. The G-fiber tends to shrink or expand in its longitudinal or transverse directions when those biomechanical processes occur. We denoted the strains of the dimensional changes of the single G-fiber in the longitudinal and the diametral directions as $\varepsilon_L$ and $\varepsilon_T$, respectively, which were simulated by the formula derived in our previous paper. Correctly speaking, it is not a model for the behaviors of an isolated fiber, since the constitutive equations used in the formulation consider the conditions of shear restraint imposed by neighboring fibers. Basic formula to calculate $\varepsilon_L$ and $\varepsilon_T$ contains several parameters, which can be categorized into a few groups as follows.

*Anatomical factors* (See Fig. 1)

$r_0, r_1, r_2, r_3$ : Outer radii in CML, the S1, the S2, and the G layers.
$r_4$ : Innermost radius in the G-fiber.
$\rho_0, \rho_1, \rho_2, \rho_3$ : Respectively, ratios of the outer radii to the inner radii in the CML, the S1, the S2, and the G layers. $\rho_0 = r_0/r_1$, $\rho_1 = r_1/r_2$, $\rho_2 = r_2/r_3$, $\rho_3 = r_3/r_4$.
$h$ : Thickness of the CML ($= r_0 - r_1$).
$\theta$ : Microfibril angle in the S2 layer (MFA).

*Mechanical factors*

$E_1, E_2, E_3$ : Young's moduli of the framework bundles of the oriented polysaccharide in the direction parallel to the molecular chain of the cellulose in the S1, the S2, and the G layers, respectively.
$S_1, S_2, S_3$ : Double shear moduli of the isotropic skeletons of the matrix substances in the S1, the S2, and the G layers, respectively.
$S_0$: Double shear modulus of the CML.

*Internal expansive terms*

$\varepsilon_1^f, \varepsilon_2^f, \varepsilon_3^f$ : Internal strains caused in the polysaccharide framework bundles in the directions parallel to the cellulose molecular chains in the S1, the S2, and the G-layers, respectively.
$\varepsilon_1^m, \varepsilon_2^m, \varepsilon_3^m$ : Internal strains caused in the matrix skeletons in the S1, the S2, and the G-layers, respectively. Those internal strains are caused by the changes of the physical state in the cell wall.

Basic equations to calculate the dimensional changes of the single G-fiber

The basic equations which gives $\varepsilon_L$ and $\varepsilon_T$ were derived as follows in our previous paper.[6)]

$$\dot{\varepsilon}_L = f_{11}(\mathbf{p})\dot{\varepsilon}_1^m + f_{12}(\mathbf{p})\dot{\varepsilon}_2^m + f_{13}(\mathbf{p})\dot{\varepsilon}_3^m + f_{14}(\mathbf{p})\dot{\varepsilon}_1^f + f_{15}(\mathbf{p})\dot{\varepsilon}_2^f + f_{16}(\mathbf{p})\dot{\varepsilon}_3^f + f_{17}(\mathbf{p})\dot{P}_L$$
$$\dot{\varepsilon}_T \left(=\dot{\varepsilon}_t|_{r=r1}\right) = f_{21}(\mathbf{p})\dot{\varepsilon}_1^m + f_{22}(\mathbf{p})\dot{\varepsilon}_2^m + f_{23}(\mathbf{p})\dot{\varepsilon}_3^m + f_{24}(\mathbf{p})\dot{\varepsilon}_1^f + f_{25}(\mathbf{p})\dot{\varepsilon}_2^f + f_{26}(\mathbf{p})\dot{\varepsilon}_3^f + f_{27}(\mathbf{p})\dot{P}_L \quad (1)$$

where a dot on each quantity stands for the derivative by an elapsed time *t*. Coefficients $f_{11}, f_{12}, \cdots, f_{27}$ are functions of **p**, and **p** is a parameter vector whose components are $\rho_0, \rho_1, \rho_2, \rho_3, \theta, E_1, E_2, E_3, S_0, S_1, S_2,$ and $S_3$. A part of those parameters depend on *t* during the cell wall maturation, or the moisture adsorption. $P_L$ stands for an axial traction which acts on both ends of the G-fiber. We can calculate the dimensional change of the single wood fiber by integrating the differential equations (1) along the physical state change of the cell wall.

Time (or moisture) dependent behaviors of the parameters

*Maturation process of the cell wall*

The amorphous constituent, such as xylan and lignin, are irreversibly accumulated among the gaps of the polysaccharide bundle after the completion of the polysaccharide framework of the cellulose microfibril (CMF) and other oriented polyose. In this process, the amorphous constituent hardens into the matrix skeleton. Thus, $S_1$, $S_2$, and $S_3$ tend to increase monotonously from very small values to their final values. Moreover, the amount of the substance increases irreversibly inside the matrix skeleton whose volume is spatially limited. As the inevitable consequence, internal strains $\varepsilon_1^m$, $\varepsilon_2^m$, $\varepsilon_3^m$ are induced in the S1, the S2, and the G-layers, respectively.

It is considered that time dependent changes in $E_1$, $E_2$, and $E_3$ are quite smaller than in $S_1$, $S_2$, and $S_3$ since the polysaccharide framework had been already completed before the matrix substance starts to deposit. However, we should not ignore a possibility that an aging effects, such as an increase in the crystallinity of the CMF, would generate internal strains $\varepsilon_1^f$, $\varepsilon_2^f$, $\varepsilon_3^f$ in the polysaccharide framework bundle.[3,6,9-11)] In such a case, we need to assume a certain value for each of them.

*Drying process in the cell wall*

Since the completed xylem (i.e. green wood) contains much water, we need to remove it before converting the wood as natural resources into the building or furniture timber. In this process, the water molecule is discharged from the absorption site in the matrix skeleton, then, the matrix skeleton tends to shrink and harden. This means that $S_1$, $S_2$, $S_3$, $\varepsilon_1^m$, $\varepsilon_2^m$, and $\varepsilon_3^m$ tend to change their values monotonously in accordance with the moisture desorption. At the same time, a certain physicochemical change may occur in the bundle of the CMF. However, it is natural to consider that changes of $E_1$, $E_2$, $E_3$ and values of $\varepsilon_1^f$, $\varepsilon_2^f$, and $\varepsilon_3^f$ are quite smaller than those of $S_1$, $S_2$, $S_3$, $\varepsilon_1^m$, $\varepsilon_2^m$, and $\varepsilon_3^m$ since the crystal domain, which is a main component of the polysaccharide framework, almost does not participate in the adsorption of the water molecule.

Determination of the values to be assumed for parameters in eqs.(1)

*Anatomical factors $\rho_0$, $\rho_1$, $\rho_2$, $\rho_3$, and $\theta$*

To determine the values of $\rho_0$, $\rho_1$, $\rho_2$, and $\rho_3$, it is required to know the ratio of the area of each layer to the total crosscut area of a single wood fiber. Then, we interrelate the parameters $\rho_0$, $\rho_1$, $\rho_2$, and $\rho_3$ by using the following formulus (see APPENDIX (A)).

$$\rho_0 \rho_1 \rho_2 = \frac{1}{\sqrt{1-s}}, \quad \rho_0 = 1 + \frac{h}{r_1}, \quad \rho_2 = \frac{1}{\rho_1 (1 + h/r_1)\sqrt{1-s}},$$

$$\rho_3 = \sqrt{\frac{1-s}{(1-s) - f \cdot g/N_g}} \quad (\text{for } N_g \neq 0), \quad \text{or} \quad \rho_3 = 1 \quad (\text{for } N_g = 0), \tag{2}$$

where $s$ and $g$ stand for the area ratios of the lignified layer (= CML+S1+S2) and the G layer in the domain of the wood fiber, respectively. $f$ and $N_g$ stand for the numbers of the wood fiber and the G-fiber per unit area in the domain of the wood fiber. Those are experimentally determined values. To determine the values of $\rho_0$, $\rho_1$, $\rho_2$ by using the eqs.(2), we need to give at least two of them. In the present calculation, with reference to the authors' previous studies,[5,12] we hypothesized 0.025 as the value of $h/r_1$, and 1.1 as the value of $\rho_1$. Thereafter, for each measuring point of the released strain, we calculated the values of $\rho_2$ and $\rho_3$ by using eqs.(2). Estimated values of $\rho_2$ and $\rho_3$ are displayed in Table 2.

Table 2. Values of the parameters $\rho_2$ and $\rho_3$ which are estimated from eqs.(2).

| Measuring positions | 1 | 2 | 3 | 4 | 5 | 6 | 7 | 8 | 9 | 10 | Mean |
|---|---|---|---|---|---|---|---|---|---|---|---|
| $\rho_2$ | 1.625 | 1.411 | 1.462 | 1.448 | 1.596 | 1.474 | 1.413 | 1.439 | 1.558 | 1.519 | 1.494[1] |
| $\rho_3$ | 1.527 | 1.598 | 1.616 | 1.645 | 1.367 | 1.274 | 1.000[2] | 1.000[2] | 1.000[2] | 1.000[2] | 1.505[3] |

1) Not significant for difference among the positions (by the test of the goodness of fit).
2) No G-fiber was formed at the positions 7, 8, 9, and 10.
3) Mean value among the 6 positions where the G-fiber is formed.

$\theta$ is one of the anatomical factors in **p**. In the present simulation, we used the measured values of the MFA in the S2 layer of the N- and the G-fibers, which are displayed in Table 1.

*Mechanical factors $E_1$, $E_2$, $E_3$, $S_1$, $S_2$, $S_3$, $S_0$*

The S1, the S2, and the G-layers can be regarded as the parallel composites of the crystalline bundle of cellulose and the matrix skeleton, then, the simple mixture law is applied to calculate the values of $E_1$, $E_2$, $E_3$, $S_1$, $S_2$, and $S_3$ as follows:[12]

$$E_1 = A_1 \times C_1 \times E_{cry}, \quad E_2 = A_2 \times C_2 \times E_{cry}, \quad E_3 = A_3 \times C_3 \times E_{cry},$$
$$S_1 = \frac{(1-A_1C_1)E_{matr}}{1+v}, \quad S_2 = \frac{(1-A_2C_2)E_{matr}}{1+v}, \quad S_3 = \frac{(1-A_3C_3)E_{matr}}{1+v}, \quad (3)$$

where $v$ is Poisson ratio, which is hypothesized to be 0.5 in the same way as our previous papers.[6,11-13] $C_1$, $C_2$, and $C_3$ are crystallinity indices of the polysaccharide framework in the S1, the S2, and the G-layers, respectively. $A_1$, $A_2$, and $A_3$ are weight ratios of the polysaccharide framework in respective layers. In this study, the values of $A_1$, $A_2$, and $A_3$ are assumed in Table 3.

$E_{matr}$ is Young's modulus of the molded matrix substance, which clearly depends on the elapsed time during the cell wall maturation (or moisture content during the moisture adsorption). With reference to Cousins's experiments,[14,15] it is assumed that $E_{matr}$= 2 GPa at the green condition, and $E_{matr}$= 4~6 GPa at the dried condition. On the other hand, it is considered that Young's modulus of the cellulose crystal along the direction parallel to the molecular chain ($E_{cry}$) is not affected by the moisture adsorption. With reference to Nishino et al's study,[16] we assume $E_{cry}$ = 134 GPa regardless of the moisture content.

Then, we assumed the values and $t$-dependent patterns of $E_1$, $E_2$, $E_3$, $S_1$, $S_2$, $S_3$ as displayed in Table 3 on the basis of the above-mentioned discussions and the after-mentioned Subsidiary Conditions, provided that the non-crystalline region in the framework bundle was regarded as the matrix substance from the mechanical point of view. The value of $S_0$ was calculated by the method described in our previous paper.[12]

Internal expansive terms $\varepsilon_1^f$, $\varepsilon_2^f$, $\varepsilon_3^f$, $\varepsilon_1^m$, $\varepsilon_2^m$, $\varepsilon_3^m$

Neither of the values nor $t$-dependent patterns can be measured for $\varepsilon_1^f$, $\varepsilon_2^f$, $\varepsilon_3^f$, $\varepsilon_1^m$, $\varepsilon_2^m$, $\varepsilon_3^m$. However, we can optimize their values and $t$-dependent patterns so as to obtain a reasonable simulation.

**Table 3.** Values of parameters used for the simulation.
(a) Chemical composition in each layer.

| Layer | Polysaccharide framework | Matrix substance | Crystallinity of the framework |
|---|---|---|---|
| CML | 15 (%) | 85 (%) | 100 (%) |
| S1 | 26 | 74 | $C_1$ |
| S2 | 52 | 48 | $C_2$ |
| G | 90 | 10 | $C_3$ |

(b) Time-dependent changes of the mechanical properties of constituents.

| Time [#] | $E_{cry}$ | $E_{matr}$ (in S1) [1)] | $E_{matr}$ (in S2) [1)] | $E_{matr}$ (in G) [1)] | $S_0$ [1)] |
|---|---|---|---|---|---|
| $t = 0 \sim T_1$ [2)] | 134 | increase from 0 to 2 | 0 | 0 | 4 |
| $t = T_1 \sim T_2$ [2)] | 134 | 2 | increase from 0 to 2 | 0 | 4 |
| $t = T_2 \sim T_3$ [2)] | 134 | 2 | 2 | increase from 0 to 2 | 4 |
| $t = T_3 \sim T_4$ [3)] | 134 | increase from 2 to 4 | increase from 2 to 4 | increase from 2 to 4 | 4 |

(c) Geometrical properties of the layers.

| Time [#] | $\rho_0$ | $\rho_1$ | $\rho_2$ | $\rho_3$ [4)] | $\theta$ [5), 6)] |
|---|---|---|---|---|---|
| $t = 0 \sim T_1$ [2)] | 1.025 | 1.1 | 1.494 | 1.0 or 1.505 | 27.2 or 23.1 |
| $t = T_1 \sim T_2$ [2)] | 1.025 | 1.1 | 1.494 | 1.0 or 1.505 | 27.2 or 23.1 |
| $t = T_2 \sim T_3$ [2)] | 1.025 | 1.1 | 1.494 | 1.0 or 1.505 | 27.2 or 23.1 |
| $t = T_3 \sim T_4$ [3)] | 1.025 | 1.1 | 1.494 | 1.0 or 1.505 | 27.2 or 23.1 |

1) Unit: GPa. 2) Maturation process after the formation of the polysaccharide framework in the S1, the S2, and the G-layers. 3) Drying process from the FSP ($t = T_3$) to the oven-dried state ($t = T_4$). $\tau = (T_4 - t)/(T_4 - T_3)$.
4) $\rho_3 = 1.0$ in the N-fiber. $\rho_3 = 1.505$ in the G-fiber.
5) Unit: deg. 6) $\theta = 27.2$ in the N-fiber. $\theta = 23.1$ in the G-fiber. # Assumed with reference to Subsidiary condition 1.

# Results

Young's modulus of the green G-layer

*Experimental results*

Matured secondary xylem of kohauchiwakaede consists of four domains of tissues, i.e. the wood fiber, the vessel element, the ray parenchyma, and the axial parenchyma. It is considered that those tissues are arranged in a row in the direction parallel to the axis of wood fiber, then, the following formula can be used for calculating the longitudinal Young's modulus of the TW xylem ($E_L^X$) by the simple law of mixture.

$$E_L^X = \frac{1}{F + V + R + P}\left(F \cdot E_L^F + V \cdot E_L^V + R \cdot E_L^R + P \cdot E_L^P\right), \quad (4)$$

where $E_L^F$, $E_L^V$, $E_L^R$, and $E_L^P$ are Young's modulus of respective tissues under the green condition, and $F+V+R+P=1$. Considering $E_L^V/E_L^F \ll 1$, $E_L^R/E_L^F \ll 1$, and $E_L^P/E_L^F \ll 1$, we obtain $E_L^F = E_L^X/F$. In the case of Kohauchiwakaede, the amount of the axial parenchyma is quite lower than that of the other tissue, and its morphological feature is almost similar as the wood fiber cell excepting the fact that the wall thickness of the axial parenchyma is more or less smaller than that of the wood fiber. In this study, for the simplification, we did not distinguish the axial parenchyma from the wood fiber when we determined the values of $F$, $V$, $R$, and $P$.

According to the observations, there was no significant difference among the measuring points on the periphery as to the morphological properties of the G-fiber, e.g. the thickness of the lignified layer, that of the G-layer, and their morphological appearance. The same can be said in the case of the N-fiber. Then, applying the simple mixture law to the fiber domain that is regarded as a parallel composite of the G-fiber and the N-fiber, we obtain the following formula:

$$E_L^F = \phi \cdot E_L^g + (1-\phi)E_L^n = (E_L^g - E_L^n)\phi + E_L^n, \quad \text{where } \phi = N_g/f, \ N_g + N_n = f, \qquad (5)$$

where $E_L^g$ and $E_L^n$ are respectively the axial Young's modulus of the green G-fiber and that of the green N-fiber, and $\phi$ is the relative frequency of the G-fiber in the fiber domain. On the other hand, we obtained the relationship between $\phi$ and $E_L^F$ (= $E_L^X/F$) as shown in Table1, which was approximated by the following linear regression:

$$E_L^F = 7.74\ \phi + 8.50 \quad (r = 0.857^{**}). \qquad (6)$$

Then, comparing eqs.(5) and (6) directly, we obtain

$$E_L^g = 16.24 \ [\text{GPa}], \quad E_L^n = 8.50 \ [\text{GPa}], \qquad (7)$$

provided that we did not use the data obtained from the measuring point 5 when deriving eq. (6) for the following reason. The observed value of the longitudinal Young's modulus at the measuring point 5 was quite larger regardless of having very small amount of G-fiber formation, therefore, estimated value of $E_L^g$ becomes abnormally larger at the position 5 than at the other positions. It is supposedly to say that something error happened when measuring the elastic modulus of the specimen at the position 5.

*Simulation using the wood fiber model*

In this simulation, we assumed the condition of the steady moisture state (green condition, i.e. the state at $t = T_3$ in Table 3). Then, every component in **p** must be constant, and both $d\varepsilon_i^m$ and $d\varepsilon_i^f$ (i = 1, 2, 3) should be all nil. Then, from eqs.(1), we obtain the following formula to calculate the longitudinal Young's modulus of the wood fiber ($E_L$):

$$E_L = \{1/(\pi r_0^2)\}dP_L/d\varepsilon_L = \{1/(\pi r_0^2)\}/f_{17}(\mathbf{p}). \qquad (8)$$

The values assumed in Table 3 were used for the simulation using eq.(8). Firstly, we optimized the values of $C_1$ and $C_2$ in eqs.(3) so as to simulate the experimentally determined value of $E_L^n$ (= 8.50GPa). In this simulation, we assumed that the degree of crystallinity in the framework bundle of the oriented polysaccharide is identical in the S1 and the S2 for convenience since there is no reason for considering that properties of the CMF are different each other between in the S1 and in the S2 layer. Thereafter, we applied the optimized values of $C_1$ and $C_2$ to the simulation of $E_L$ in the green G-fiber, and optimized the values of $C_3$ so as to obtain the experimentally determined value of $E_L^g$ (= 16.24GPa). Finally, the optimized values of $C_1$, $C_2$, and $C_3$ became:

$$C_1 \ (= C_2\ ) = 0.494, \quad C_3 = 0.221. \qquad (9)$$

From this result, we calculate the longitudinal Young's modulus of the lignified layer in the N-fiber ($E_N^n$), that of the lignified layer in the G-fiber ($E_N^g$), and that of the G-layer ($E_G^g$) as follows:

$$\text{In the N-fiber}: E_N^n = 13.13 \text{ [GPa]}$$

$$\text{In the G-fiber}: E_N^g = 16.28 \text{ [GPa]}, \quad E_G^g = 28.27 \text{ [GPa]}. \tag{10}$$

Growth strain in the G-layer

*Experimental Results*

The wood fiber, the vessel element, the ray parenchyma, and the axial parenchyma in the differenciating xylem tend to deform during their secondary wall maturation. Thus, the growth strain is generated in the maturing xylem. Infinitesimal increase in the longitudinal growth strain of the xylem at the macroscopic level ($\varepsilon_L^X$) can be expressed as the following formula by the simple mixture law:

$$d\varepsilon_L^X = \frac{F \cdot E_L^F \cdot d\varepsilon_L^F + V \cdot E_L^V \cdot d\varepsilon_L^V + R \cdot E_L^R \cdot d\varepsilon_L^R + P \cdot E_L^P \cdot d\varepsilon_L^P}{F \cdot E_L^F + V \cdot E_L^V + R \cdot E_L^R + P \cdot E_L^P}. \tag{11}$$

Where $d\varepsilon_L^F$, $d\varepsilon_L^V$, $d\varepsilon_L^R$, and $d\varepsilon_L^P$ are infinitesimal increses of the longitudinal growth strain in respective tissues. Assuming $E_L^V/E_L^F \ll 1$, $E_L^R/E_L^F \ll 1$, $E_L^P/E_L^F \ll 1$, and $F+V+R+P=1$, we obtain

$$d\varepsilon_L^X \cong d\varepsilon_L^F.$$

Moreover, we obtain the following formula:

$$d\varepsilon_L^X (\cong d\varepsilon_L^F) = \frac{\phi \cdot E_L^g \cdot d\varepsilon_L^g + (1-\phi) \cdot E_L^n \cdot d\varepsilon_L^n}{\phi \cdot E_L^g + (1-\phi) \cdot E_L^n} = \frac{\left(E_L^g \cdot d\varepsilon_L^g - E_L^n \cdot d\varepsilon_L^n\right) \cdot \phi + E_L^n \cdot d\varepsilon_L^n}{\left(E_L^g - E_L^n\right) \cdot \phi + E_L^n} \tag{12}$$

where $d\varepsilon_L^g$ and $d\varepsilon_L^n$ are respective increments in the longitudinal growth strain of the G-fiber and that of the N-fiber respectively. By integrating the eq.(12) along the cell wall maturation, we can obtain the growth strain of the newly-formed xylem ($\varepsilon_L^X$).

In order to integrate the eq.(12), we also need to know the changes of $E_L^g$ and $E_L^n$ during the process of the secondary wall maturation. It is considered that deposition of the matrix constituents have almost no effect on the increases of $E_L^g$ and $E_L^n$ since the stiffness of the matrix substance is quite smaller than that of the framework bundle. Therefore, it is rather natural to consider that increases of $E_L^g$ and $E_L^n$ are caused by a certain qualitative change of the CMF, such as further crystallization of cellulose.[17] Unfortunately, it is still quite difficult to know the time-dependent change of the CMF crystallinity in the cell wall. In the present study, for convenience, we assumed that the crystallinity in each layer is almost unchanged during the cell wall maturation, then, we hypothesized the $E_L^g$ and $E_L^n$ becomes constant through the cell wall maturation.

The growth stress generation is a biomechanical process during the maturation (lignification) of the secondary wall.[11,17,18] Thus, we integrate eq.(12) along the cell wall maturation in the G-fiber. As the result, we obtain the following formula:

$$\varepsilon_L^X (\cong \varepsilon_L^F) = \frac{\phi \cdot E_L^g \cdot \varepsilon_L^g + (1-\phi) \cdot E_L^n \cdot \varepsilon_L^n}{\phi \cdot E_L^g + (1-\phi) \cdot E_L^n} = \frac{(E_L^g \cdot \varepsilon_L^g - E_L^n \cdot \varepsilon_L^n) \cdot \phi + E_L^n \cdot \varepsilon_L^n}{(E_L^g - E_L^n) \cdot \phi + E_L^n} \tag{13}$$

where $\varepsilon_L^g = \int_{\text{Maturation process}} d\varepsilon_L^g$, $\varepsilon_L^n = \int_{\text{Maturation process}} d\varepsilon_L^n$.

Results (7) were used as the values of $E_L^g$ and $E_L^n$ in this formula. Observed relationship between $\phi$ and $\varepsilon_L^X$ (=$\varepsilon_L^F$), which was shown in Table 1, was approximated by the following curvilinear regression:

$$\varepsilon_L^F = -0.5554 + \frac{0.6003}{\phi + 1.098} \quad (r = 0.956^{***}) \tag{14}$$

Then, comparing the eqs.(13) and (14) directly, we obtained the growth strains of G-fiber ($\varepsilon_L^g$) and the N-fiber ($\varepsilon_L^n$) as follows:

$$\varepsilon_L^g = -0.2693 \, [\%], \quad \varepsilon_L^n = -0.0087 \, [\%]. \tag{15}$$

*Simulation based on the G-fiber model*

We integrated the basic formula (1) during the G-fiber wall maturation under the assumption of $dP_L=0$. As initial conditions, we adopted $\varepsilon_L(t)|_{t=0} = 0$, $\varepsilon_T(t)|_{t=0} = 0$. Results (9) were used as the values of $C_1$, $C_2$, and $C_3$. Values of the parameters assumed in Table 3 were also used for the calculation. Then, we optimized the increments and $t$-dependent patterns of $\varepsilon_1^f$, $\varepsilon_2^f$, $\varepsilon_3^f$, $\varepsilon_1^m$, $\varepsilon_2^m$, and $\varepsilon_3^m$ so as to obtain the results (15). However, before integrating eqs.(1), we need to know how the maturation of the G-fiber wall proceeds.

Some scientists clarified the lignification process in the secondary wall of the softwood tracheid and the hardwood normal-fiber,[19,20] on the other hand, maturation of the G-fiber has remained still unclear. Lately, based on the technique of immuno-TEM observation, Kim et al. discovered that the activity of the peroxidase is localized in the secondary wall rather after the completion of the G-layer.[21] This suggests that lignification proceeds in the secondary wall after the formation of the thick G-layer. Then, with reference to those investigations, we assumed the following conditions as to the maturation of the G-fiber.

Subsidiary Condition 1: Firstly, lignification in the S1 layer starts at $t = 0$ after the formation of the frameworks of the cellulose and the other oriented polysaccharide in the secondary wall and the G-layer, and ends at $t = T_1$. This is the first integration interval. Secondly, the lignification in the S2 layer starts at $t = T_1$, and ends at $t = T_2$. This is the second integration interval. In the G-layer, deposition of a certain matrix substance should proceed, however, no lignification occurs. In this study, as the third integration interval, the deposition of the matrix substance in the G-layer starts at $t = T_2$ and ends at $t = T_3$. Then the G-fiber maturation is completed at $t = T_3$. $S_1$, $S_2$, and $S_3$ tend to increase monotonously and smoothly from very small values to their final values in their respective integration intervals.

Then, we integrate eqs.(1) as follows:

$$\varepsilon_L \left(= \int_{t=0}^{t=T_3} \left(\frac{d\varepsilon_L}{dt}\right) dt\right) = \underbrace{\int \left(\frac{d\varepsilon_L}{dt}\right) dt}_{\text{The first integration interval}} + \underbrace{\int \left(\frac{d\varepsilon_L}{dt}\right) dt}_{\text{The second integration interval}} + \underbrace{\int \left(\frac{d\varepsilon_L}{dt}\right) dt}_{\text{The third integration interval}} \quad (1')$$

We need to impose certain subsidiary conditions on values and *t*-dependent patterns of $\varepsilon_1^f$, $\varepsilon_2^f$, $\varepsilon_3^f$, $\varepsilon_1^m$, $\varepsilon_2^m$, and $\varepsilon_3^m$ so as to simulate the observed values of $\varepsilon_L^n$ and $\varepsilon_L^g$. By the way, in the case of the softwood xylem, the observed relationship between the longitudinal growth strain and the MFA in the latewood tracheid can be simulated by supposing [increment in $\varepsilon_1^m$] = 1%, [increment in $\varepsilon_2^m$] = 0.5%, and [increment in $\varepsilon_1^f$] = [increment in $\varepsilon_2^f$] = -0.15%.[11] With reference to this result, we assumed the following subsidiary conditions.

Subsidiary Condition 2: The values of $\varepsilon_1^m$ and $\varepsilon_2^m$ take positive values. Each of them increases monotonously and smoothly from 0 to a certain value (= *increment*) as the lignification proceeds in each integration interval.[11] It is natural to consider that increments in $\varepsilon_1^m$ and $\varepsilon_2^m$ depend on the lignin content in respective layers. This is based on *the lignin swelling hypothesis*. However, we assume $\varepsilon_3^m = 0$, since no lignification occurs in the G-layer. On the other hand, the values of $\varepsilon_1^f$ and $\varepsilon_2^f$ take negative values. Each of them tends to change monotonously and smoothly from 0 to a certain value (= *increment*) with the maturation in each integration interval. This postulates *the cellulose tension hypothesis* which considers that the CMF framework tends to contract in the direction parallel to the cellulose molecular chain with the aging of the CMF.[17, 22]

Firstly, we simulated the generation of the growth strain of the N-fiber ($\varepsilon_L^n = -0.0087\%$) by integrating eq.(1') under the above subsidiary conditions, and optimized the increments in $\varepsilon_1^f$ and $\varepsilon_2^f$ so as to obtain the observed value of $\varepsilon_L^n$ (= -0.0087%). Thereafter, we tried to simulate the generation of the growth strain of the G-fiber ($\varepsilon_L^g = -0.2693\%$) and optimized the increment $\varepsilon_3^f$. In this simulation, we assumed the following subsidiary condition in addition to above two conditions:

Subsidiary Condition 3: According to the observations by using the light- or ultraviolet microscopes, there is no specific difference in the morphological appearance between the secondary wall of the N-fiber and that of the G-fiber.[23] From this fact, we assumed that *t*-dependent patterns and increments in each of $\varepsilon_1^m$, $\varepsilon_2^m$, $\varepsilon_1^f$ and $\varepsilon_2^f$ take identical values between in the N-fiber and in the G-fiber.

$S_1$, $\varepsilon_1^f$, and $\varepsilon_1^m$ are all expressed as monotonously increasing (or decreasing) functions of *t* in the first integration interval. $S_2$, $\varepsilon_2^f$, and $\varepsilon_2^m$ are also monotonously increasing (or decreasing) functions of *t* in the second integration interval. The same can be said for $S_3$, $\varepsilon_3^f$, and $\varepsilon_3^m$ in the third integral interval. Each of those monotonously increasing (decreasing) functions can be transformed into the function which do not contain $T_1$, $T_2$, and $T_3$ explicitly by transforming the integral variable

$t$ into $\gamma$ ($=t/T_1$; $0<t<T_1$), or $\xi$ ($=(t-T_1)/(T_2-T_1)$; $T_1<t<T_2$) or $\kappa$ ($=(t-T_2)/(T_3-T_2)$; $T_2<t<T_3$). Moreover, we know those variable transformations alter corresponding integration intervals in eq.(1') into an identical one that is from 0 to 1. Thus, the concrete value of eq.(1') does not depend on $T_1$, $T_2$, and $T_3$. Furthermore, we should notice that integration of eq.(1') is not affected by the functional shapes of $t$-dependent variables if each variable would change its value very smoothly in each integration interval. This is rather reasonable since we consider that the $t$-dependent changes of those variables gradually proceed by the maturation of the matrix skeleton in respective layers (see APPENDIX (B)).

Thus, we can optimize the value of the increment in $\varepsilon_3^f$ as displayed in Table 5 which became quite larger than those in $\varepsilon_1^f$ and $\varepsilon_2^f$ as shown in Tables 4.

**Table 4.** Combination on the increments in $\varepsilon_1^m$, $\varepsilon_2^m$, $\varepsilon_1^f$, $\varepsilon_2^f$ which optimize the observed value of $\varepsilon_L^n$. (Units : %)

| $\varepsilon_1^m$ (at $t=T_1$) 1) | 1 | 0.8 | 0.6 | 0.4 | 0.2 | 0 | -0.2 | -0.4 |
|---|---|---|---|---|---|---|---|---|
| $\varepsilon_2^m$ (at $t=T_2$) 1) | 0.5 | 0.4 | 0.3 | 0.2 | 0.1 | 0 | -0.1 | -0.2 |
| $\varepsilon_1^f$ (at $t=T_1$) 2) (= $\varepsilon_2^f$ at $t=T_2$) | -0.178 | -0.145 | -0.112 | -0.078 | -0.045 | -0.012 | 0.021 | 0.055 |
| $\varepsilon_T^n$ (at $t=T_2$) 3) | 0.170 | 0.135 | 0.099 | 0.063 | 0.028 | -0.0078 | -0.043 | -0.079 |

1) Assumed arbitrarily under Subsidiary Conditions 1 and 2.
2) Calculated from each pair of $\varepsilon_1^m$ and $\varepsilon_2^m$ so as to give the observed value of $\varepsilon_L^n$.
3) Calculated from each combination of $\varepsilon_1^m$, $\varepsilon_2^m$, $\varepsilon_1^f$, $\varepsilon_2^f$.

**Table 5.** Estimated value of the increment in $\varepsilon_3^f$ which gives the observed value of $\varepsilon_L^g$. (Units : %)

| $\varepsilon_1^m$ (at $t=T_1$) # | 1 | 0.8 | 0.6 | 0.4 | 0.2 | 0 | -0.2 | -0.4 |
|---|---|---|---|---|---|---|---|---|
| $\varepsilon_2^m$ (at $t=T_2$) # | 0.5 | 0.4 | 0.3 | 0.2 | 0.1 | 0 | -0.1 | -0.2 |
| $\varepsilon_1^f$ (at $t=T_1$) ## (= $\varepsilon_2^f$ at $t=T_2$) | -0.178 | -0.145 | -0.112 | -0.078 | -0.045 | -0.012 | 0.021 | 0.055 |
| $\varepsilon_3^f$ (at $t=T_3$) ### | -0.721 | -0.737 | -0.754 | -0.770 | -0.786 | -0.802 | -0.818 | -0.834 |
| $\varepsilon_T^g$ (at $t=T_3$) #### | 0.333 | 0.298 | 0.263 | 0.228 | 0.193 | 0.158 | 0.123 | 0.088 |

\# Assumed arbitrarily under Subsidiary Conditions 1, 2, and 3.
\#\# Calculated for each pair of $\varepsilon_1^m$ and $\varepsilon_2^m$ so as to obtain the observed value of $\varepsilon_L^n$.
\#\#\# Calculated for each combination of $\varepsilon_1^m$, $\varepsilon_2^m$, $\varepsilon_1^f$, $\varepsilon_2^f$ so as to obtain the observed value of $\varepsilon_L^g$.
\#\#\#\# Calculated for each combination of $\varepsilon_1^m$, $\varepsilon_2^m$, $\varepsilon_1^f$, $\varepsilon_2^f$, and $\varepsilon_3^f$.

Drying shrinkage of the G-layer

*Experimental results*

We can describe the shrinking process of the wood as the function of the moisture content $\tau$ that is normalized by the moisture content at the fiber saturation point (FSP). We denote the longitudinal shrinking process of the wood as $\alpha_L^X(\tau)$. According to the definition, the longitudinal shrinkage $\alpha_L^X(\tau)$ must satisfy the following boundary condition, $\alpha_L^X(\tau)\big|_{\tau=1} = 0$.
$\alpha_L^X(\tau)|_{\tau=0}\ (=\alpha_L^X)$ means the oven-dried shrinkage of the wood. An infinitesimal increase of the moisture content ($d\tau$) causes an infinitesimal change in the shrinkage of the wood ($d\alpha_L^X$), which is described as the following formula:

$$d\alpha_L^X = \frac{F \cdot \overline{E}_L^F \cdot d\alpha_L^F + V \cdot \overline{E}_L^V \cdot d\alpha_L^V + R \cdot \overline{E}_L^R \cdot d\alpha_L^R + P \cdot \overline{E}_L^P \cdot d\alpha_L^P}{F \cdot \overline{E}_L^F + V \cdot \overline{E}_L^V + R \cdot \overline{E}_L^R + P \cdot \overline{E}_L^P}, \qquad (16)$$

where $d\alpha_L^F$, $d\alpha_L^V$, $d\alpha_L^R$, and $d\alpha_L^P$ stand for infinitesimal changes of the longitudinal shrinkage in respective tissues. $\overline{E}_L^F, \overline{E}_L^V, \overline{E}_L^R$, and $\overline{E}_L^P$ are respective Young's moduli at the moisture content ). Assuming $\overline{E}_L^V/\overline{E}_L^F \ll 1$, $\overline{E}_L^R/\overline{E}_L^F \ll 1$, $\overline{E}_L^P/\overline{E}_L^F \ll 1$, and $F+V+R+P=1$, we obtain

$$d\alpha_L^X \cong d\alpha_L^F.$$

We apply the simple mixture law to the fiber domain consisting of the N- and G-fibers in parallel, then, we obtain the following formula:

$$d\alpha_L^F (\cong d\alpha_L^X) = \frac{\phi \cdot \overline{E}_L^g \cdot d\alpha_L^g + (1-\phi) \cdot \overline{E}_L^n \cdot d\alpha_L^n}{\phi \cdot \overline{E}_L^g + (1-\phi) \cdot \overline{E}_L^n} = \frac{(\overline{E}_L^g \cdot d\alpha_L^g - \overline{E}_L^n \cdot d\alpha_L^n) \cdot \phi + \overline{E}_L^n \cdot d\alpha_L^n}{(\overline{E}_L^g - \overline{E}_L^n) \cdot \phi + \overline{E}_L^n}, \quad (17)$$

where $d\alpha_L^g$ and $d\alpha_L^n$ are infinitesimal changes in the shrinkage of the G-fiber and that of the N-fiber, respectively, and $\overline{E}_L^g$ and $\overline{E}_L^n$ are axial Young's moduli of G-fibers and N-fiber, respectively.

We obtain an oven-dried shrinkage of the wood fiber domain $\alpha_L^F\ (=\alpha_L^X(\tau)|_{\tau=0})$ by integrating eq.(17) from an arbitrary $\tau$ to FSP ($\tau=1$) and extrapolating $\tau \to 0$, provided that we need to know the $\tau$-dependent patterns of $\overline{E}_L^g$ and $\overline{E}_L^n$ in advance. Then, we tentatively expressed $\overline{E}_L^g$ and $\overline{E}_L^n$ as follows:

$$\overline{E}_L^n = E_L^n \cdot \xi(\tau), \qquad \overline{E}_L^g = E_L^g \cdot \zeta(\tau), \qquad (18)$$

where $\xi(\tau)$ and $\zeta(\tau)$ are monotonously decreasing functions for $\tau$, and they satisfy $\xi(\tau)|_{\tau=1}=1$, and $\zeta(\tau)|_{\tau=1}=1$. $E_L^n$ and $E_L^g$ are constants, which stand for the axial Young's moduli of the green N-fiber and the green G-fiber, respectively. For simplification, we assumed $\xi(\tau) = \zeta(\tau)$ for all $\tau$, which means the decreasing pattern of the longitudinal Young's modulus in the G-fiber is similar as that in the N-fiber. Then, eq.(15) becomes

$$d\alpha_L^F (\cong d\alpha_L^X) = \frac{(E_L^g \cdot d\alpha_L^g - E_L^n \cdot d\alpha_L^n) \cdot \phi + E_L^n \cdot d\alpha_L^n}{(E_L^g - E_L^n) \cdot \phi + E_L^n}. \qquad (19)$$

Under those assumptions, we substituted the results (7) to $E_L^g$ and $E_L^n$ in eq.(19). As the initial conditions, $\alpha_L^g(\tau)|_{\tau=1}=\alpha_L^n(\tau)|_{\tau=1}=0$, were required. Thus, eq.(19) can be integrated during the increasing process of the moisture content (from an arbitrary $\tau$ to $\tau=1$). We obtain the oven-dried shrinkage of the wood fiber domain $\alpha_L^F$ ( $=\alpha_L^F(\tau)|_{\tau=0}$) as the following formula.

$$\alpha_L^X (\cong \alpha_L^F) = \frac{(E_L^g \cdot \alpha_L^g - E_L^n \cdot \alpha_L^n) \cdot \phi + E_L^n \cdot \alpha_L^n}{(E_L^g - E_L^n) \cdot \phi + E_L^n} \quad . \tag{20}$$

Observed relationship between $\phi$ and $\alpha_L^F$, which was shown in Table 1, was approximated by the following curvilinear regression:

$$\alpha_L^F = -2.429 + \frac{2.363}{\phi + 1.098} \quad . \qquad (r = 0.867^{***}) \tag{21}$$

Then, comparing eqs.(20) and (21) directly, we obtain the oven-dried shrinkage of the N-fiber ($\alpha_L^n$) and the G-fiber ($\alpha_L^g$) as follows:

$$\alpha_L^n = 0.2771 \ [\%], \qquad \alpha_T^g = 1.3026 \ [\%] \ . \tag{22}$$

*Simulation based on the G-fiber model*

Free dimensional change of the single wood fiber due to the moisture adsorption was simulated on the basis of the conditions assumed in Table 3. Thus, $dP_L$ should be null in eqs.(1). For convenience, we calculated the swelling deformation of the wood fiber model $\varepsilon_L(\tau)$ by integrating eqs.(1) from $\tau=0$ to $\tau=1$. Relationship between the swelling $\varepsilon_L$ ($=\varepsilon_L(\tau)|_{\tau=1}$) and the oven-dried shrinkage $\alpha_L$ ($=\alpha_L(\tau)|_{\tau=0}$) are related each other by the following formulas:

$$\alpha_L = \frac{\varepsilon_L}{\varepsilon_L + 1} \quad , \qquad \varepsilon_L = \frac{\alpha_L}{1 - \alpha_L} \quad . \tag{23}$$

The integral interval for calculating $\varepsilon_L$ ($=\varepsilon_L(\tau)|_{\tau=1}$) is from the oven-dried state ($t=T_4$; $\tau=0$) to the fiber saturation point ($t=T_3$; $\tau=1$). It is regarded that increasing moisture content $\tau$ is equivalent to the reciprocal elapsed time $t$. The results (9) were used as the values of $C_1$, $C_2$, and $C_3$ in this simulation. Then, we optimized the increments in $\varepsilon_1^m$, $\varepsilon_2^m$, $\varepsilon_3^m$, $\varepsilon_1^f$, $\varepsilon_2^f$, and $\varepsilon_3^f$ so as to obtain the observed values of $\alpha_L^n$ and $\alpha_L^g$ .

Swelling of the softwood tracheid cell wall is mainly caused by the swelling of the matrix substance, e.g. hemicellulose, lignin, and noncrystalline cellulose.[12,24-27] Therefore, it is quite natural to postulate that $\varepsilon_1^m$, $\varepsilon_2^m$, and $\varepsilon_3^m$ take positive values with the increase of the moisture content, and increase monotonously from 0 to the final values, that is to say, increments.

Firstly, we simulated the swelling of the N-fiber ($\varepsilon_L^n$ = 0.2779%). Concretely to say, we optimized the values of increments in $\varepsilon_1^m$, $\varepsilon_2^m$, $\varepsilon_1^f$, and $\varepsilon_2^f$ so as to give the observed value of the oven-dried shrinkage $\alpha_L^n$ (= 0.2771%). In the present simulation, we assumed $\varepsilon_1^m = \varepsilon_2^m = \varepsilon_3^m$, and $\varepsilon_1^f = \varepsilon_2^f$ for convenience.

Optimized values of the increments in $\varepsilon_1^m$, $\varepsilon_2^m$, $\varepsilon_1^f$, and $\varepsilon_2^f$ were obtained by the simulation as displayed in Table 6. In our previous report, we succeeded in simulating the observed relationships between the longitudinal and the tangential swellings, and the MFA in the clear wood specimen of sugi (*Cryptomeria japonica*) by supposing that $\varepsilon_1^m = \varepsilon_2^m = 12 \sim 15\%$, and $\varepsilon_1^f = \varepsilon_2^f = 0\sim 1\%$.[12] In the present simulation, optimized $\varepsilon_1^f$ and $\varepsilon_2^f$ became very small but negative, which means that the polysaccharide framework bundles in the S1 and the S2 layers tend to contract in the direction parallel to the cellulose molecular chains in spite that the moisture content increases in the cell wall. This gives us a very strange impression. It is impossible for the authors to give any comment on this result at this stage, then, we withhold our mention on this result for the time being. However, their absolute values were so small as compared with the increment in $\varepsilon_1^m$ and $\varepsilon_2^m$.

Secondly, we simulated the oven-dried shrinkage of the G-fiber ($\alpha_L^g = 1.3026\%$), and optimized the value of the increment in $\varepsilon_3^f$. In this simulation it is assumed that increment in each of $\varepsilon_1^m$, $\varepsilon_2^m$, $\varepsilon_1^f$, and $\varepsilon_2^f$ takes an identical value between in the N-fiber and in the G-fiber (see *Subsidiary Condition 3*). For convenience, we assume $\varepsilon_1^m = \varepsilon_2^m = \varepsilon_3^m$ in this simulation. Thereafter, we optimized the value of the increment in $\varepsilon_3^f$ so as to obtain the observed value of $\alpha_L^g$. Results are displayed in Table 7. Optimized value of the increment in $\varepsilon_3^f$ became a large positive value which is quite different from those in $\varepsilon_1^f$ and $\varepsilon_2^f$.

**Table 6.** Combinations of the increments of $\varepsilon_1^m$, $\varepsilon_2^m$, $\varepsilon_1^f$, $\varepsilon_2^f$ which optimize the observed value of $\alpha_L^n$. (Units: %)

| | | | | | | | | |
|---|---|---|---|---|---|---|---|---|
| $\varepsilon_1^m (= \varepsilon_2^m)$ at $t = T_4$ [1) 2)] | 15 | 12 | 9 | 6 | 3 | 0 | -3 | -6 |
| $\varepsilon_1^f (= \varepsilon_2^f)$ at $t = T_4$ [2) 3)] | -3.345 | -2.605 | -1.865 | -1.126 | -0.386 | 0.353 | 1.094 | 1.833 |
| $\alpha_T^n$ [4) 5)] | 9.017 | 7.361 | 5.643 | 3.861 | 2.010 | 0.086 | -1.915 | -3.997 |

1) Assumed arbitrarily.
2) Positive sign means that each cell wall constituent swells during water sorption.
3) Calculated from each pair of $\varepsilon_1^m$ and $\varepsilon_2^m$ so as to optimize the observed value of $\alpha_L^n$.
4) Calculated from each combination of $\varepsilon_1^m$, $\varepsilon_2^m$, $\varepsilon_1^f$, $\varepsilon_2^f$.
5) Positive sign means that the wood fiber shrinks in diameter from the FSP to the oven-dried state.

**Table 7.** Estimated value of the increments in $\varepsilon_3^f$ which gives the observed value of $\alpha_L^g$. (Units : %)

| | | | | | | | | |
|---|---|---|---|---|---|---|---|---|
| $\varepsilon_1^m (= \varepsilon_2^m = \varepsilon_3^m)$ at $t = T_4$ [#, *] | 15 | 12 | 9 | 6 | 3 | 0 | -3 | -6 |
| $\varepsilon_1^f (= \varepsilon_2^f)$ at $t = T_4$ [##, *] | -3.345 | -2.605 | -1.865 | -1.126 | -0.386 | 0.353 | 1.094 | 1.833 |
| $\varepsilon_3^f$ (at $t = T_4$) [###, *] | 5.078 | 4.782 | 4.486 | 4.190 | 3.897 | 3.603 | 3.304 | 3.010 |
| $\alpha_T^g$ [####, **] | 13.66 | 11.14 | 8.461 | 5.616 | 2.588 | -0.641 | -4.089 | -7.784 |

\# Assumed arbitrarily.
\#\# Calculated from each pair of $\varepsilon_1^m$ and $\varepsilon_2^m$ so as to optimize the observed value of $\alpha_L^n$.
\#\#\# Calculated from each combination of $\varepsilon_1^m$, $\varepsilon_2^m$, $\varepsilon_3^m$, $\varepsilon_1^f$, $\varepsilon_2^f$ so as to obtain the observed value of $\alpha_L^g$.
\#\#\#\# Calculated from each combination of $\varepsilon_1^m$, $\varepsilon_2^m$, $\varepsilon_3^m$, $\varepsilon_1^f$, $\varepsilon_2^f$, and $\varepsilon_3^f$.
\* Positive sign means that each cell wall constituent swells during water sorption.
\*\* Positive sign means that the wood fiber shrinks in diameter from the FSP to the oven-dried state.

## Discussion

Young's modulus of the green G-layer ($E_G^g$)

According to the results (7), (9), and (10), the predicted Young's modulus of the green G-layer ($E_G^g$) became 2.15 times as large as that of the lignified layer in N-fiber ($E_N^n$), and 1.74 times as large as the one in the G-fiber ($E_N^g$). In any case, we can say that the longitudinal Young's modulus of the G-layer becomes more or less larger than that of the lignified layer in the G- and N-fibers. By the way, the predicted value of Young's modulus of the lignified layer in the G-fiber ($E_N^g$) became slightly larger than the one in the N-fiber ($E_N^n$). This is because we calculated the value of $E_N^g$ in due consideration of an experimental fact that the MFA of the S2 layer in the G-fiber was a little smaller than in the N-fiber. This may be one of the factors to increase the Young's modulus of the TW xylem.

It is well known that the TW becomes very stiffer in the longitudinal direction as it dries. The increase of Young's modulus of the TW xylem due to drying is highly correlated with the percentage of the G-fiber in the fiber domain.[4] This suggests that the G-layer becomes abruptly rigid as the water molecule is released. However, the propriety of this suggestion remains to be proved in a next work.

Strangely to say, predicted value of the relative crystallinity in the framework bundle of the oriented polysaccharide in the G-layer was quite smaller than that in the secondary wall (see results (9)). According to the formula derived in our previous paper,[12] Young's modulus of the G-layer is highly dependent on the ratio of the cellulosic component. In the present simulation, we supposed it to be 90% in the G-layer, which may be a little larger than the true value in the green G-layer. It is imagined that the G-layer contains not a few amount of non-crystalline polyose, e.g. hemicellulose. As another possibility, we indicate a fact that the green G-layer is highly swollen by the water, which causes an apparent decrease in the relative crystallinity of the cellulose in the green G-layer. Hitherto, we have referred to Norberg and Meier's classical data on the chemical and physical properties of the G-layer in aspen.[7] However, we need to verify their conclusion critically for various species.

Growth strain in the G-layer ($\varepsilon_3^f$)

Simulated value of $\varepsilon_3^f$ is quite larger negative than that in the lignified layer ($\varepsilon_1^f$ and $\varepsilon_2^f$). This indicates that a large contractive internal strain originates in the polysaccharide framework of the G-layer in the direction of the cellulose molecular chain, which causes a high longitudinal tensile growth stress in the TW xylem.

Shrinkage and swelling of the G-layer due to moisture adsorption

Many authors have considered that the polysaccharide framework does not swell or shrink by the moisture adsorption. However, present simulation shows that the value of $\varepsilon_3^f$, which is the swelling ability of the polysaccharide framework in the G-layer, becomes a large positive value. Conversely to say, the polysaccharide framework in the G-layer tends to shrink in the direction parallel to the cellulose molecular chain during the moisture adsorption. This means that the high longitudinal drying shrinkage in the TW xylem is induced by the drying shrinkage of the G-layer in its axial direction. Lately, Clair and Thibaut observed that the dried G-layer tends to be depressed from surrounding lignified layer by using the SEM observation,[28] which supports the predicted results in the present simulation.

## APPENDIX (A)   Deriving eqs. (2).

We denote the number of the G-fiber in the wood fiber domain with an area of $A$ as $G$ and that of the N-fiber as $N$, provided that $G + N = X$. We set the following assumption.

(Assumption A) The thickness of the lignified wall in the G-fiber is identical with that of the N-fiber regardless of the measuring position.

This assumption is not so inappropriate to the wood fiber domain in the real xylem since the observed values of $s$ and $X/A$ $(= f)$ became almost unchanged regardless of measuring positions as seen from Table 1. Moreover, we set the following assumptions.

(Assumption B) The diameter of the G-fiber is similar as that of the N-fiber.

(Assumption C) Cellular arrangement in the crosscut surface of the xylem takes a tessellation structure consisting of a polygonal cell.

Then, we can connect the area ratio of the lignified layer [$s$], and that of the gelatinous layer [$g$] to $\rho_0$, $\rho_1$, $\rho_2$, and $\rho_3$ under the assumptions (A), (B), and (C).

It may be a little hasty to apply calculated values of $\rho_0$, $\rho_1$, $\rho_2$, and $\rho_3$ to the simulation using eqs.(1) since the crosscut shape of the G-fiber model displayed in Fig.1 is accurately circular. However, we know that the hexagon is the most closely allied to the circle in shape among the polygons which constitute the tessellation arrangement. Then, we set the following assumption

(Assumption D) Crosscut shape of the wood fiber in the cellular arrangement is a hexagon with an area of $(3\sqrt{3}/2)r_0^2$ as displayed in Fig.2.

We denote the thickness of the lignified layer as $(\sqrt{3}/2)(r_0 - r_3)$, and that of the G-layer as $(\sqrt{3}/2)(r_3 - r_4)$. Distances from the central point of the hexagonal to the lignified and the gelatinous layers are denoted as $(\sqrt{3}/2)r_3$ and $(\sqrt{3}/2)r_4$, respectively. $s$ can be given as the following formula:

$$s = \frac{3}{2}\sqrt{3}(r_0^2 - r_3^2)X/A = 1 - \frac{3}{2}\sqrt{3} \cdot r_3^2 \cdot f . \tag{A1}$$

In a similar manner, $g$ is given as the following formula:

$$g = \frac{3}{2}\sqrt{3}(r_3^2 - r_4^2)G/A = (1-s)\frac{G}{X} - \frac{3}{2}\sqrt{3} \cdot r_4^2 \cdot \frac{G}{A} . \tag{A2}$$

$X/A$, $G/A$, $s$, and $g$ can be decided experimentally as displayed in Table 1. Then, from (A1) and (A2), we can obtain $r_3$ and $r_4$ as follows:

$$r_3 = \frac{1}{3}\sqrt{2\sqrt{3}(1-s)\frac{A}{X}} \quad , \quad r_4 = \frac{1}{3}\sqrt{2\sqrt{3}\left((1-s)\frac{A}{X} - g\frac{A}{G}\right)} \quad (G \neq 0). \tag{A3}$$

If we denote $r_0/r_1$, $r_1/r_2$, $r_2/r_3$, $r_3/r_4$ as $\rho_0$, $\rho_1$, $\rho_2$, $\rho_3$ respectively, we obtain the following equation:

$$\rho_0 \rho_1 \rho_2 = \frac{1}{\sqrt{1-s}}, \quad \rho_3 = \sqrt{\frac{(1-s)}{(1-s) - g \cdot f/N_g}} \quad \text{(for } N_g \neq 0\text{)}, \quad \rho_3 = 1 \quad \text{(for } N_g = 0\text{)}, \tag{A4}$$

where $f = X/A$, and $N_g = G/A$.

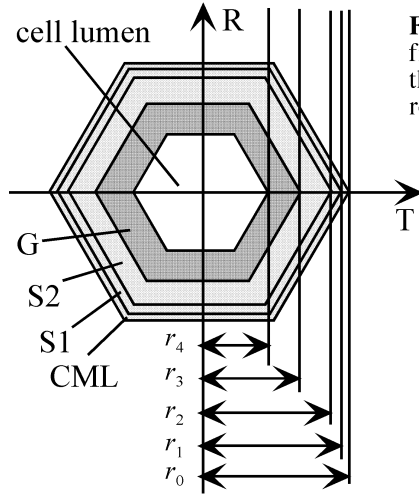

**Fig.2.** A hexagonal shaped model of the wood fiber for calculating eqs.(2). T and R stand for the tangential and the radial directions, respectively.

## APPENDIX (B) Integration (1') is not affected by the functional shapes of $t$-dependent variables, $S_1$, $S_2$, $S_3$, $\varepsilon_1^m$, $\varepsilon_2^m$, $\varepsilon_3^m$, $\varepsilon_1^f$, $\varepsilon_2^f$, and $\varepsilon_3^f$.

We introduce functions $\varphi_1$, $\varphi_2$, and $\varphi_3$ which vary from 0 to 1 in the range of $0 \leq t \leq T_3$ as follows:

$$\varphi_1(t) = \begin{cases} P(t) & (0 \leq t \leq T_1) \\ 1 & (T_1 \leq t \leq T_3) \end{cases}, \quad \varphi_2(t) = \begin{cases} 0 & (0 \leq t \leq T_1) \\ Q(t) & (T_1 \leq t \leq T_2) \\ 1 & (T_2 \leq t \leq T_3) \end{cases}, \quad \varphi_3(t) = \begin{cases} 0 & (0 \leq t \leq T_2) \\ R(t) & (T_2 \leq t \leq T_3) \end{cases}, \tag{B1}$$

where $P(t)$, $Q(t)$, and $R(t)$ are monotonously increasing and differentiable functions which vary from 0 to 1 smoothly in respective domains. With reference to Subsidiary Conditions 1 and 2, we assume the following condition as the functional shapes of $t$-dependent variables $S_i(t)$, $\varepsilon_i^m(t)$, and $\varepsilon_i^f(t)$ (i =1, 2, 3):

$$S_i(t) = k_i \cdot \varphi_i(t), \quad \varepsilon_i^m(t) = m_i \cdot \varphi_i(t), \quad \varepsilon_i^f(t) = n_i \cdot \varphi_i(t). \quad (i = 1, 2, 3) \tag{B2}$$

where $k_i$, $m_i$, and $n_i$ are constants. It is enough natural to assume this condition if those $t$-dependent variables change the values smoothly during the maturation of the matrix skeleton in their respective integration intervals. Then, by substituting (B2) into eq.(1'), we obtain the following expression:

$$\varepsilon_L = \int_0^{T_1} g_1(\varphi_1(t))\left(\frac{d\varphi_1(t)}{dt}\right)dt + \int_{T_1}^{T_2} g_2(\varphi_2(t))\left(\frac{d\varphi_2(t)}{dt}\right)dt + \int_{T_2}^{T_3} g_3(\varphi_3(t))\left(\frac{d\varphi_3(t)}{dt}\right)dt. \tag{B3}$$

where 
$$\begin{cases} g_1(\varphi_1(t)) = m_1 \cdot f_{11}(\mathbf{p})|_{S_1 = k_1 \cdot \varphi_1(t)} + n_1 f_{14}(\mathbf{p})|_{S_1 = k_1 \cdot \varphi_1(t)} & (0 \leq t \leq T_1) \\ g_2(\varphi_2(t)) = m_2 \cdot f_{12}(\mathbf{p})|_{S_2 = k_2 \cdot \varphi_2(t)} + n_2 f_{15}(\mathbf{p})|_{S_2 = k_2 \cdot \varphi_2(t)} & (T_1 \leq t \leq T_2) \\ g_3(\varphi_3(t)) = m_3 \cdot f_{13}(\mathbf{p})|_{S_3 = k_3 \cdot \varphi_3(t)} + n_3 f_{16}(\mathbf{p})|_{S_3 = k_3 \cdot \varphi_3(t)} & (T_2 \leq t \leq T_3) \end{cases}$$

From (B1), eq.(B3) is modified into the following expression.

$$\varepsilon_L = \int_{t=0}^{t=T_1} g_1(P(t))\,dP(t) + \int_{t=T_1}^{t=T_2} g_2(Q(t))\,dQ(t) + \int_{t=T_2}^{t=T_3} g_3(R(t))\,dR(t) . \tag{B4}$$

By the way, $P(t)$, $Q(t)$, and $R(t)$ change the values from 0 to 1 monotonously and continuously for elapsed time $t$ in their respective integration intervals, then, we can rewrite eq.(B4) as the following expression:

$$\varepsilon_L = \int_0^1 g_1(P)\,dP + \int_0^1 g_2(Q)\,dQ + \int_0^1 g_3(R)\,dR . \tag{B5}$$

This result indicates that the integration value in eq.(B5) does not depend on the concrete values of $T_1$, $T_2$ and $T_3$, furthermore, it is not affected by the functional shapes of $t$-dependent variables $S_1$, $S_2$, $S_3$, $\varepsilon_1^m$, $\varepsilon_2^m$, $\varepsilon_3^m$, $\varepsilon_1^f$, $\varepsilon_2^f$, and $\varepsilon_3^f$ if we assume the condition (B2).